# Laboratory Setup for Testing Low-Frequency Disturbances of Power Quality

Piotr Kuwałek
*Institute of Electrical Engineering and Electronics*
*Poznań University of Technology*
Poznań, Poland
piotr.kuwalek@put.poznan.pl

Grzegorz Wiczyński
*Institute of Electrical Engineering and Electronics*
*Poznań University of Technology*
Poznań, Poland
grzegorz.wiczynski@put.poznan.pl

*Abstract*—Low-frequency disturbances of power quality are one of the most common disturbances in the power grid. These disturbances are most often the result of the impact of power electronic and energy-saving devices, the number of which is increasing significantly in the power grid. Due to the simultaneous operation of various types of loads in the power grid, various types of simultaneous disturbances of power quality occur, such as voltage fluctuations and distortions. Therefore, there is a need to analyze this type of simultaneous interaction. For this purpose, a special and complementary laboratory setup has been prepared, which allows for the examination of actual states occurring in modern power networks. Selected research results are presented for this laboratory setup, which determine its basic properties. Possible applications and possibilities of the laboratory setup are presented from the point of view of current challenges.

## I. Introduction

Low-frequency disturbances of power quality are one of the most common disturbances in the power grid [1,2]. Low-frequency power quality disturbances include: voltage fluctuations, voltage distortions caused by higher harmonics or sub-, inter- or supra-harmonics, frequency fluctuations, etc. These disturbances occur most often as a result of the impact of electronic and energy-saving devices [3-7], the number of which in the power grid has been increasing significantly in recent years. Due to the simultaneous operation of various types of loads in the power grid, various types of simultaneous disturbances of power quality occur. One of the most common simultaneous disturbances in the low-voltage network are, for example, voltage fluctuations and voltage distortions caused by higher harmonics [1,8-10]. The "clipped cosine" voltage distortion is common in low-voltage networks [11,12] and is caused by the input stages of switching power supplies. If a load that periodically changes its operating state is connected to such a network, simultaneous voltage fluctuations and voltage distortions occur as a result [13]. In recent years, simultaneous power quality disturbances have also been shown to cause negative effects that are unobservable for individual disturbances occurring separately [13]. For example, for voltage fluctuations alone, a flicker can occur as a result of a change in the operating state at a frequency of $3f_c$, where $f_c$ is the power frequency [14]. In turn, in the case of simultaneous voltage fluctuations and voltage distortions, a flicker can occur for the load that changes its operating state with a frequency whose limit value depends on the level of supply voltage distortion and can be greater than $3f_c$ [13,15]. Therefore, it is a need to analyze this type of simultaneous interaction [16-20].

It is worth noting that in the currently applicable normative documents in the field of methods for measurement and assessment of low-frequency disturbances (e.g. the standard IEC 61000-4-30 [21] referring to the standard IEC 61000-4-7 [22] in the field of voltage/current distortion measurements or the standard IEC 61000-4-15 [23] in the field of flicker assessment (voltage fluctuations)), idealized states are considered or certain models are proposed that recreate the state of occurrence of a low-frequency disturbance of only one type. This analysis facilitates the testing of measuring instruments and simplifies the assessment of the propagation of disturbances in the power grid. It is worth noting, however, that such an analysis does not guarantee a small measurement error in the event of actual disturbance states typical of modern power networks. Examples of discrepancies in measurement results in the research with the normative approach and in the research with the approach recreating actual disturbance states can be observed for AMI smart energy meters equipped with the power quality assessment functionality [24,25]. To some extent, the simultaneous occurrence of low-frequency disturbances can be recreated using a high-power broadband arbitrary generator. However, the assessment of the propagation of disturbances in the event of their simultaneous occurrence requires a specialized laboratory setup. Taking into account the indicated needs, a special and complementary laboratory setup was prepared, which allows for the examination of the actual conditions that occur in modern power networks. The paper discusses the construction of this unique laboratory setup. Selected research results are presented for this laboratory setup, which determine its basic properties. Possible applications and possibilities of the laboratory setup are presented from the point of view of current challenges.

## II. Description of the Laboratory Setup

The specialized laboratory setup is shown in the block diagram in Fig. 1.

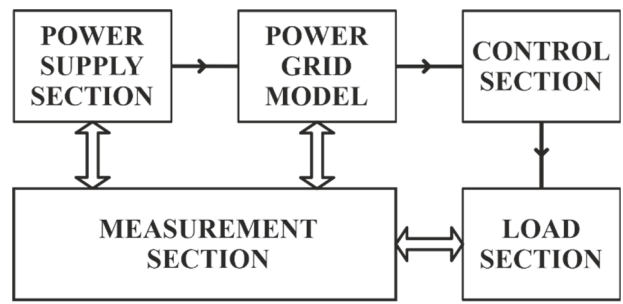

Fig. 1. The block diagram of the laboratory setup for testing low-frequency disturbances of power quality

This work was funded by National Science Centre, Poland – 2021/41/N/ST7/00397. For the purpose of Open Access, the author has applied a CC–BY public copyright licence to any Author Accepted Manuscript (AAM) version arising from this submission.

In the diagram shown in Fig. 1, five blocks are indicated:

- single-sided 3-phase power grid model with branching radial topology with branching;
- power supply section;
- control section;
- load section;
- measurement section.

The actual view of the laboratory setup is shown in Fig. 2. The presented laboratory setup can be used, for example, for:

- research that exceeds normative approaches [26,27];
- research in the field of evaluating the propagation of power quality in the power grid [28,29];
- research on the identification and localization in sources of disturbances of power quality (including, for example, sources of voltage fluctuations) [30,31];
- assessment of the interaction between power quality disturbances and other connected loads [32,33];
- post-factum examination of disturbance states in the power grid [34];
- post-factum testing of voltage-supplied loads in the presence of various power quality disturbances [35].

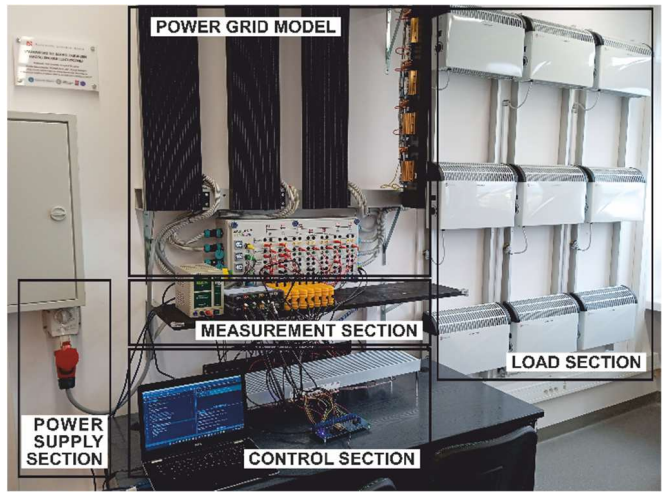

Fig. 2. The actual view of the laboratory setup

*A. Power grid model*

The most important element of the laboratory setup is a single-sided power supply model of branching 3-phase power grid with a radial topology. This is a typical model for low-voltage networks, where low-frequency disturbances of power quality occur most often. The diagram of the power grid model is shown in Fig. 3. The prepared power grid model consists of six sections, where sections III and IV are included in the first branch, and sections V and VI are included in the second branch. In the prepared power grid model, individual resistance values $R_i$ and inductance $L_i$ were selected in such a way as to obtain typical conditions typical for an overhead line. The nominal values of the resistance $R_N$ of the resistors used and the inductance $L_N$ of the coils used are presented in Table I.

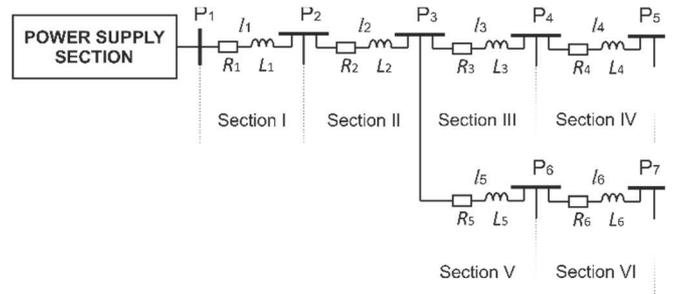

Fig. 3. The diagram of the power grid model

The individual power supply points $P_i$ of the power grid model are output to the dashboard as follows:

- power connections allowing for the connection of a load with a maximum load capacity of 32 A [copper wire with a cross-section of 4 mm$^2$];
- measurement connection allowing for the recording of voltage signals from individual points of the power grid model [LiY-CY shielded copper wire with a cross-section of 0.25 mm$^2$];
- measurement connection allowing for the recording of current signals using clamp current transformers from individual points of the power grid model [copper wire with a cross-section of 4 mm$^2$].

TABLE I. LIST OF NOMINAL VALUES OF PARAMETERS OF ELEMENTS USED IN THE POWER GRID MODEL

| Section of Power Grid Model | $R_N$ [mΩ] | $L_N$ [µH] |
|---|---|---|
| Section I | 150.0 | 100.0 |
| Section II | 150.0 | 100.0 |
| Section III | 100.0 | 220.0 |
| Section IV | 50.0 | 6.8 |
| Section V | 100.0 | 220.0 |
| Section VI | 50.0 | 6.8 |

The power grid model has lumped parameters corresponding to the distributed parameters of the real long line. Taking into account this fact and the fact that the connections for recording voltage signals are shielded, the prepared laboratory setup ensures effective suppression of external interferences that could affect the correctness of the analysis. Fig. 4 shows a photo of a part of the neutral line mounted on the radiator. It is also worth noting that it is possible to slightly modify the prepared power grid model by connecting capacitances to the power terminals in parallel (in the case of the designed model, the capacitances can be in the order of nF), which allow the model to be changed from an overhead line model to a cable line model.

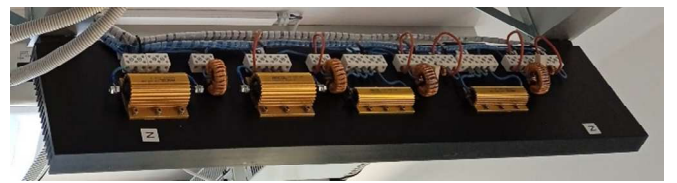

Fig. 4. The view of a fragment of the neutral line mounted on the radiator

## B. Power supply section

The power supply section ensures that the supply voltage is supplied to the power supply point $P_1$ of the prepared power grid model. The prepared power grid model allows the free connection of various types of power supply according to demand. The power supply section can include:

- supply voltage directly from the FLUKE 5500A calibrator [impact analysis for nominal conditions - functional test signals generated with an inaccuracy of ±0.03%];

- supply voltage from the cascade: Rigol DG1022z arbitrary generator - CHROMA 61502 power amplifier [analysis of impacts for conditions recreating real disturbance states in modern power networks - arbitrary signals generated in the frequency range up to 2400 Hz with an inaccuracy of 0.2% - the bandwidth depends on the power amplifier used];

- supply voltage directly from the MV/LV transformer station with a power of 630 kVA [impact analysis for real conditions - study of the propagation of disturbances generated by the tested real loads, study of the mutual influence between the actual supply voltage and the operation of the selected load in controlled conditions in which the added loads are known at the level of the power grid model];

- supply voltage from any source that is not currently available in the laboratory and which allow testing of conditions specified by a specific user.

## C. Control section

The control section allows for independent switching of connected loads, synchronously or asynchronously. The control is implemented by an SSR system based on a MOSFET transistor, which allows for achievement of maximum switching at the kHz level. In addition, the use of an SSR system based on a MOSFET transistor allows the load to be switched on and off at any time, without delays of μs between the appropriate edge of the control signal and the change of the switch state. Fig. 5 shows a simplified diagram of a single switch. Fig. 6 shows the view of individual switches mounted on the radiator.

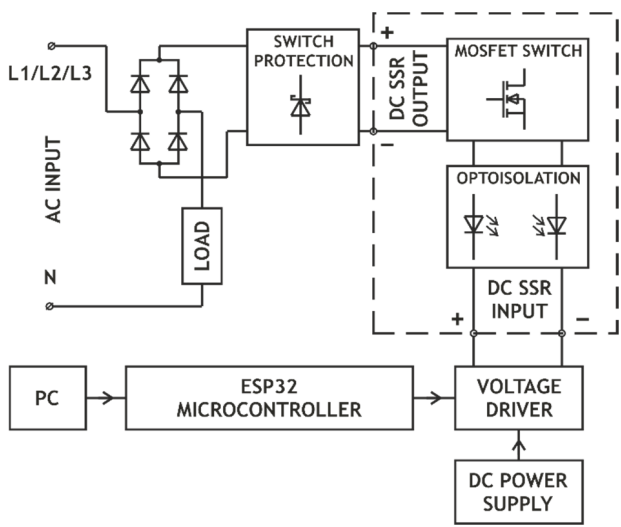

Fig. 5. The simplified diagram of a single switch

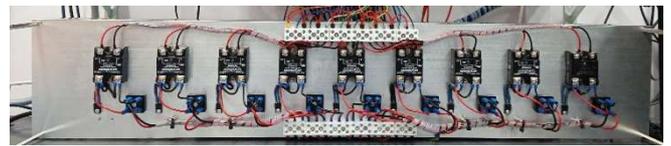

Fig. 6. The view of individual switches mounted on the radiator

The control system presented in Fig. 5, apart from the assessment of the propagation of disturbances after turning on or off a specific load, also allows for recreation of the operation of a group of loads or selected power electronic systems. In such a case, it is sufficient to use elements such as a resistor, coil, and capacitor in a series or parallel configuration as a load. The indicated elements allow for simulation of the nature of the load, and the switching frequency allows for recreation of the variability of the operating state of a specific device (e.g., inverter). The control system is operated from a computer via the ESP32 microcontroller. The individual digital outputs of the ESP32 system are fed to a voltage driver, which appropriately amplifies the signal and has an appropriate current carrying capacity. Switching of the control system on the AC side is possible by using a Greatz bridge. Additionally, the Schottky diode to the SSR output provides additional protection of the switch against possible overvoltages that can arise as a result of switching.

## D. Load section

The load section available at the laboratory setup includes:

- resistors in the form of convection heating systems with an active power of 0.75 kW, 1.25 kW, or 2 kW;

- capacitors with a capacity of 9.6 μF and a reactive power of 0.83 kvar;

- chokes with an inductance of 1.123 H and a reactive power of 0.15 kvar;

- Greatz rectifier bridges;

- selected specific loads, such as UPS with an apparent power of 1.8 kVA, electric vehicle charger, selected LED or fluorescent light sources, switching power supplies, etc.

The laboratory station is configured in such a way that it is also possible to connect any other real load supplied from the low-voltage network to analyze the disturbances emitted by it and the impact of other disturbances of power quality on the operation of this device.

## E. Measurement section

The measurement section allows for synchronous (simultaneous) or asynchronous recording of low-voltage signals with a value not exceeding 20 V. PicoScope 5444D PC oscilloscopes are used as recorders. The PicoScope 5444D PC oscilloscope provides recording with a sampling rate of 1GSa/s at 8-bit resolution or 62.5 MSa/s at 16-bit resolution. The bandwidth of this recorder is 200 MHz. The following are used to acquire voltage signals from the laboratory setup:

- Pico TA041 active differential probes, which enable recording voltages up to 700 V in the 25 MHz bandwidth and inaccuracies of ±2% [probe ratio equal to 100:1 or 10:1];

- HV301GB hall effect voltage transducers, which enable recording voltages up to 900 V in the

50 kHz bandwidth and with an inaccuracy of ±1% [converter ratio equal to 120:1].

The following are used to acquire current signals from the laboratory station:

- split-core transformer current transformers SCT013-005, which enable recording AC currents in the range of up to 5 A in the 1 kHz bandwidth [voltage signal at the transformer output of 200 mV/A possible via the built-in measurement resistance];
- Tektronix A622 hall effect clamp current transformers, which enable recording DC and AC currents in the range of up to 100 A in the bandwidth up to 100 kHz [voltage signal at the output 10 mV/A and 100 mV/A].

In addition to acquiring instantaneous values of voltage and current signals, it is possible to record selected quantities in the field of power quality using class A power quality analyzers PQ BOX 100 and PQ BOX 300, and using smart energy meters AMI. Selected values in the field of power quality include: active power $P$, reactive power $Q$, active energy $E_P$, reactive energy $E_Q$, fundamental frequency $f_c$, rms value of voltage $U$ and current $I$, index of total distortion of voltage THD$U$ and current THD$I$, short-term $P_{st}$ and long-term $P_{lt}$ flicker indicator, etc. The values recorded by the indicated meters can be used as reference indications in the case of normative manner. On the other hand, in the case of research that exceeds the normative requirements (research carried out from a scientific point of view), it is possible to develop new measures based on the recorded values of instantaneous voltages and currents, and then it is possible to compare them with the values currently recorded for the purposes of assessment of power quality. In the case of the previous sections, the measurement section can be expanded with any measuring and recording devices that carry out measurements in accordance with the required procedure from the point of view of the selected research.

## III. RESEARCH ON THE PARAMETERS OF THE PREPARED POWER GRID MODEL

To ensure the correctness of scientific research conducted in the prepared laboratory setup, it is necessary to determine the actual values of the parameters defining the prepared power grid model. Hence, the HM8118 Rohde&Schwarz HAMEG RCL bridge was used to measure the resistance $R_i$ and reactance $X_i$ of the resulting RCL series system of the prepared power grid model. The measurement inaccuracy of the RCL bridge used is ±0.05%. During the measurements, the resulting values of resistance $R_i$ and reactance $X_i$ of the series model were recorded in the frequency range from 20 Hz to 200 kHz. For each frequency, six measurements were made, on the basis of which the average values of individual parameters were determined. Measurements were made in a four-wire configuration. Table II shows the results for a frequency of 50 Hz and compares them with the nominal values. Figs. 7-10 show the test results obtained for every analyzed frequency and for individual phases at the ends of two branches (resultant value of resistance and reactance from sections I-II-III-IV and from sections I-II-V-VI). The characteristics of individual lines overlap and can be difficult to distinguish. The trend of the obtained results also coincides with other configurations of individual sections, therefore, taking into account the limitation of the number of pages, the presentation of these measurement results is omitted.

TABLE II. LIST OF ELEMENTS NOMINAL VALUES OF PARAMETERS AND DATA OBTAINED FROM MEASUREMENTS FOR ELEMENTS USED IN THE POWER GRID MODEL

| | Section | I | II | III | IV | V | VI |
|---|---|---|---|---|---|---|---|
| $R_i$ [m$\Omega$] | Nominal | **150.0** | **150.0** | **100.0** | **50.0** | **100.0** | **50.0** |
| | L1 | 172.6 | 163.4 | 126.8 | 61.0 | 129.6 | 59.8 |
| | L2 | 179.1 | 164.7 | 125.5 | 60.8 | 131.2 | 60.9 |
| | L3 | 176.3 | 169.4 | 125.7 | 62.6 | 130.7 | 61.3 |
| | N | 181.5 | 164.3 | 129.2 | 61.1 | 127.3 | 62.5 |
| $X_i$ [m$\Omega$] | Nominal | **31.4** | **31.4** | **69.1** | **2.1** | **69.1** | **2.1** |
| | L1 | 32.9 | 33.7 | 68.8 | 2.7 | 69.2 | 2.7 |
| | L2 | 33.4 | 33.3 | 68.9 | 2.7 | 68.9 | 2.8 |
| | L3 | 33.6 | 33.3 | 67.5 | 2.7 | 67.7 | 2.7 |
| | N | 33.5 | 33.1 | 68.8 | 2.6 | 68.2 | 2.6 |

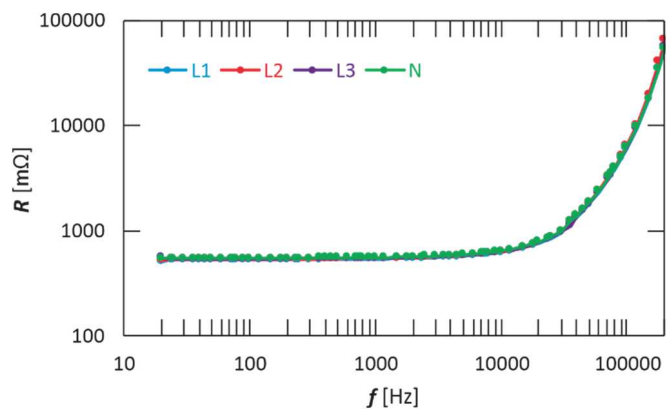

Fig. 7. The average value of the resultant resistance measured for sections I-II-III-IV

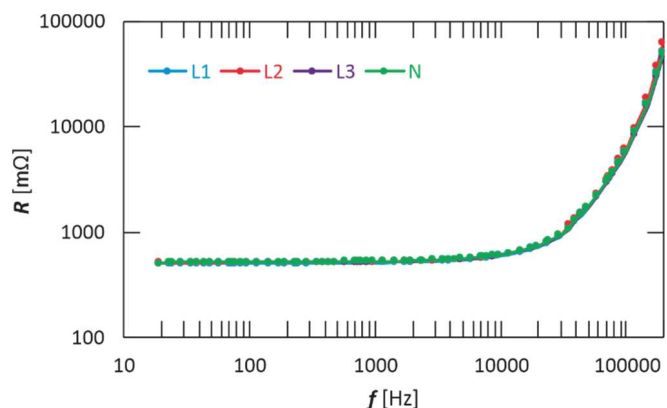

Fig. 8. The average value of the resultant resistance measured for sections I-II-V-VI

Analyzing the results presented in Table II, some discrepancies can be noticed in relation to the nominal values of the elements placed on the heat sink. The discrepancies result, among others, from the technological dispersion of the elements made, as well as from the connections of the indicated elements with individual sockets located on the dashboard. Each additional contact (screwed or soldered) increases the resulting resistance value of the indicated model. Therefore, greater discrepancies are observed for the resultant resistance value than for the resultant reactance value, the discrepancies of which are mainly related to the extension of individual connections (parasitic inductance increases). Analyzing the results presented in Figs. 7-8, a significant increase in resistance at high frequencies can be noticed, which, among others, is associated with an increase in the share of parasitic elements of the resistors used and the skin effect. Analyzing the results presented in Figs. 9-10, it can be noticed a quasi-linear increase in reactance with increasing frequency, which is consistent with expectations, because with a constant inductance value, reactance increases linearly with increasing frequency. For high frequency values, a slight nonlinearity is observed, which can also be the result of a parasitic share of capacitance in the designed system.

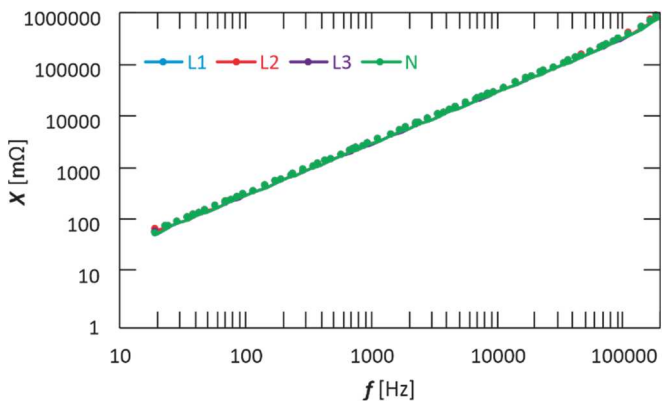

Fig. 9. The average value of the resultant reactance measured for sections I-II-III-IV

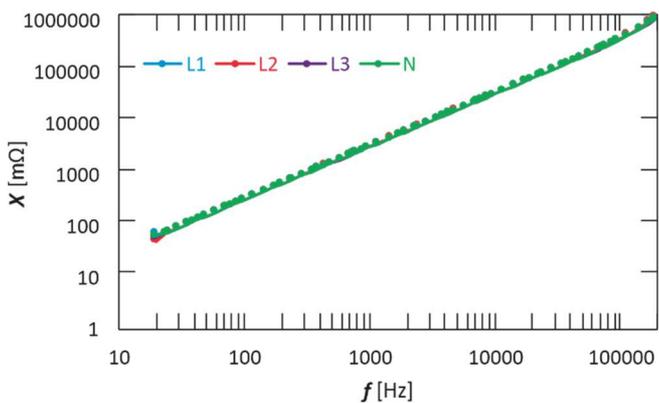

Fig. 10. The average value of the resultant reactance measured for sections I-II-V-VI

## IV. Conclusion

The paper presents a special and complementary laboratory setup that enables the examination of the actual states that occur in modern power networks. The paper discusses the construction of this unique laboratory setup. Selected research results are presented for this laboratory setup, which determine its basic properties. Possible applications and possibilities of the prepared laboratory stand are presented from the point of view of current challenges. The presented stand enables the analysis of simultaneous low-frequency disturbances of power quality. Moreover, the laboratory setup enables the assessment of the propagation of selected power quality disturbances and enables the testing of various types of currently used loads or groups of these loads. The laboratory setup also enables research, including validation, of various techniques for identification and localization o sources of power quality disturbances in conditions recreating the actual conditions occurring in modern power grids. In addition to the basic infrastructure cooperating with the prepared power grid model, it is possible to expand the laboratory setup with infrastructure important from the point of view of any other research, which can be used in the future for new directions of research.